\documentclass[onecolumn,epjc3]{svjour3}
\usepackage{mathtext,bbm,amsmath,amsfonts,amssymb,indentfirst,syntonly,graphicx}
\smartqed  
\RequirePackage{graphicx}
\RequirePackage{subfigure}
\RequirePackage{xcolor}
\def\bc{\begin{center}}
\def\ec{\end{center}}
\def\be{\begin{eqnarray}}
\def\ee{\end{eqnarray}}

\journalname{Eur. Phys. J. C}

\begin{document}
\title{Finslerian MOND versus the Strong Gravitational Lensing of the Early-type Galaxies}
\author{Zhe Chang \thanksref{addr1,addr2},
           Ming-Hua Li \thanksref{addr1},
           Xin Li \thanksref{addr1,addr2},
           Hai-Nan Lin \thanksref{e1,addr1},
           Sai Wang \thanksref{addr1}.
}
\thankstext{e1}{e-mail: linhn@ihep.ac.cn}
\institute{Institute of High Energy Physics, Chinese Academy of Sciences, 100049 Beijing, China\label{addr1}\and
           Theoretical Physics Center for Science Facilities, Chinese Academy of Sciences, 100049 Beijing, China\label{addr2}
}
\date{Received: date / Accepted: date}
\maketitle

\begin{abstract}
The gravitational lensing of Bullet Clusters and early-type galaxies pose serious challenges on the validity of MOND. Recently, Finslerian MOND, a generalization of MOND in the framework of Finsler gravity, has been proposed to explain the mass discrepancy problem of Bullet Cluster 1E 0657\---558. In this paper, we check the validity of the Finslerian MOND in describing the strong gravitational lensing of early-type galaxies. The investigation on ten strong lenses of the CASTLES samples shows that there is no strong evidence for the existence of dark matter.
\end{abstract}

\section{Introduction}

The modified Newtonian dynamics (MOND) \cite{Milgrom1983a,Milgrom1983b}, or its relativistic counterpart tensor-vector-scalar (TeVeS) theory \cite{Bekenstein2004}, has achieved great success in solving the mass discrepancy problem of galaxy rotation curves. According to MOND, the Newtonian dynamics is modified as $\mu(\vec{a}/a_0)\vec{a}=-\nabla \Phi_N$, where $\vec{a}$ is the acceleration, $\Phi_N$ is the Newtonian potential and $a_0$ is a critical acceleration. The interpolation function $\mu(x)$ is a positive, smooth, and monotonically increasing function which approaches $x$ when $\mid x \mid \ll 1$ and equals unity when $x$ goes to infinity. In the Newton range $a \gg a_0$, the dynamics reduces to that of Newton. In the deep-MOND range $a\ll a_0$, the rotational velocity becomes asymptotically flat $v_{\infty}=(GMa_0)^{1/4}$. The interpolation function $\mu(x)$ smoothly joints the two ranges. With only one universal parameter, MOND can well explain the observed rotation curves of spiral galaxies \cite{Sanders1996,Sanders:1998gr}. However, the weak and strong gravitational lensing of the early-type galaxies and the Bullet Clusters pose serious challenges to MOND \cite{Ferreras2005,Ferreras2008,Ferreras2009,Ferreras2012,Angus2006,Angus2007}.

The first challenge comes from the strong gravitational lensing of the early-type galaxies. Zhao et al. \cite{Zhao2006} analyzed 18 strong gravitational lenses from the CfA-Arizona Space Telescope Lens Survey (CASTLES) and concluded that MOND can do a successful fit. However, the mass-to-light ratios derived in their paper show a high fluctuation. Some lenses have extremely high mass-to-light ratios, while others require negative dark matter components. The later works of Ferreras et al. \cite{Ferreras2005,Ferreras2008,Ferreras2009,Ferreras2012} showed that a significant amount of dark matter is still required  even in the framework of MOND in order to match the observations. One of the reasons for the different conclusions is that they derived the photometric mass from different methods. The predicted gravitational mass in MOND depends on the choice of the interpolation function $\mu(x)$. Although the dark matter may diminish if one choose a more slowly increasing interpolation function, it will conflict with the data of galaxy rotation curves \cite{Ferreras2008}.

Another challenge arises from the weak and strong gravitational lensing of the Bullet Clusters. The Bullet Cluster 1E 0657\---558 is one of the most well studied Bullet Clusters at present. The X-ray survey of this cluster shows a lot of anomalies \cite{Clowe2004,Bradac:2006er,Brownstein:2007sr}. The mass profile of the Bullet Cluster has a sub-peak, which is a few kpc away from the main peak. The gravitational center deviates from the photometric center. The gravitational mass is much higher than the photometric mass. The features of the Bullet Cluster 1E 0657\---558 indicate that the spacetime may be anisotropic. As an isotropic theory, MOND cannot reconcile the observed $\Sigma$-map (which is reconstructed from photometric data) with the convergence $\kappa$-map (which is reconstructed from gravitational lensing) of the Bullet Cluster 1E 0657\---558 \cite{Angus2006,Angus2007}.

Up to now, the most successful way to solve these problems is the dark matter hypothesis. In the dark matter scenario, our universe contains a significant number of non-luminous mass. It is dark matter that fills the gap between the dynamical mass and the photometric mass. Unfortunately, after decades of heavy research, no direct evidences of the existence of dark matter have been found. Since dark matter neither emits nor absorbs electromagnetic radiations, it is difficult be detected directly. The only hints that the dark matter may exist come from astrophysics and cosmology. We know only a little about the dark matter on the particle physics level. The interaction between dark matter and ordinary matter is not clear. Although some experiments are claimed to have found the dark matter \cite{Adriani2009,Aguilar2013}, all of these arise wide controversies. This motivates us to search for other explanations of the discrepancy between the dynamical mass and the photometric mass.

In a very recent paper, Li et al. proposed a generalization of MOND in the framework of Finsler gravity, the so called Finslerian MOND, to explain the observed convergence $\kappa$-map of Bullet Cluster 1E 0657\---558 \cite{Li2013}. The metric in the Finslerian MOND is very like Schwarzschild's metric, except that the spatial distance $r$ is replaced by a new distance $R=rf\big{(}v(r)\big{)}$, where $f\big{(}v(r)\big{)}$ is the Finslerian rescaling factor which depends on the Finslerian structure of the spacetime. Choosing a special form of $f\big{(}v(r)\big{)}$, the Finslerian MOND may reduce to the famous MOND of Milgrom. By adding the contributions of dipole and quadrupole effects, it was showed that the Finslerian MOND can well reproduce the main properties of the observed convergence $\kappa$-map of Bullet Cluster 1E0657-558. Later, Chang et al. used the Finslerian MOND to fit the rotation curve data of spiral galaxies, and pointed out that the flatness of the galaxy rotation curves may relate to the anisotropy of the spacetime \cite{Chang2013}.

In this paper, we will check whether the Finslerian MOND can solve the mass discrepancy problems of strong gravitational lensing of the early-type galaxies. The rest of the paper is organized as follows: In section \ref{sec:FMOND}, we give a brief introduction to the Finslerian MOND, and derive the gravitational deflection angle of light in the framework of Finslerian MOND. In section \ref{sec:lensing}, we use ten strong gravitational lensing early-type galaxies from the CASTLES sample to check the validity of the Finslerian MOND. It is found that the Finslerian MOND can match the observations within uncertainties. Finally, discussion and conclusion are given in section \ref{sec:discussion}.

\section{Finslerian MOND and the deflection of light}\label{sec:FMOND}

Finsler geometry is a generalization of Riemann geometry. Instead of defining an inner product on the manifold in Riemann geometry, Finsler geometry is based on the so called Finsler structure $F$, which is defined on the bundle and has the property $F(x, \lambda y) \equiv \lambda F(x, y)$ for all $\lambda > 0$. Here, $x$ is the position coordinate, $y\equiv dx/d\tau$ is the velocity, and $\tau$ is an affine parameter such as the proper time. The second order partial differentiation of the Finsler structure $F$ with respect to the velocity $y$ gives the Finsler metric \cite{Bao2000}
\begin{equation}
  g_{\mu\nu}\equiv\frac{\partial}{\partial y^{\mu}}\frac{\partial}{\partial y^{\nu}}\left(\frac{1}{2}F^2\right)\,.
\end{equation}
The arc length in the $n$-dimensional Finsler spacetime is given as
\begin{equation}
  \int_s^r F\left(x^1,\cdots,x^n;\frac{dx^1}{d\tau},\cdots,\frac{dx^n}{d\tau}\right)d\tau\,.
\end{equation}

In the post-Newtonian approximation, the law of gravity in Finsler spacetime is very similar to that in Newtonian case. The only difference is that the spatial distance in Minkowski space $r$ is now replaced by that in Finsler space $R$ \cite{Li2013}. The Finsler structure (metric) takes the form
\begin{equation}\label{eq:finsler-metric}
  F^2d\tau^2=\left(1-\frac{2GM}{R}\right)d\tau^2-\left(1-\frac{2GM}{R}\right)^{-1}dR^2-R^2(d\theta^2+\sin^2\theta d\phi^2).
\end{equation}
The metric in Eq.(\ref{eq:finsler-metric}) is very like Schwarzschild metric. Therefore, the dynamics of a test particle moving in the gravitational potential of a point particle of mass $M$ reads
\begin{equation}\label{eq:finsler-gravity}
  \frac{GM}{R^2}=\frac{v^2}{R}\,,
\end{equation}
where $G$ is the Newtonian gravitational constant and $v$ is the rotational velocity  of the testing particle. The spatial distances in the two different spacetimes (that of Finsler and of Minkowski) are related by \cite{Li2013}
\begin{equation}
  R=rf\big{(}v(r)\big{)}\,,
\end{equation}
where the Finslerian rescaling factor $f\big{(}v(r)\big{)}$, which depends on the Finsler structure $F$, is a function of velocity $v$ and position coordinate $r$. Thus, the gravitational force between two particles depends not only their positions, but also their velocities. Theoretically, the explicit form of $f\big{(}v(r)\big{)}$ can be derived by solving the field equations. Each Finsler structure $F$ corresponds to one form of  $f\big{(}v(r)\big{)}$. Unfortunately, the Finsler structure is defined on the tangent bundle of the manifold, and we do not know how to write the energy-momentum tensor on the bundle at present. What we can do is to assume one appropriate form of $f\big{(}v(r)\big{)}$ and check its validity in describing the astrophysical phenomena. Details of the Finsler gravity can be found in references \cite{Li2013,Chang2008,Chang2009,Li2010a,Li2010b}.

The Finsler gravity equation (\ref{eq:finsler-gravity}) includes Milgrom's MOND as a special case. If the Finsler structure $F$ takes the form
\begin{equation}
f\big{(}v(r)\big{)}=\sqrt{1-\left(\frac{GMa_0}{v^4}\right)^2}\,,
\end{equation}
then Eq.(\ref{eq:finsler-gravity}) becomes
\begin{equation}\label{eq:finsler1}
  \frac{GM}{r^2}=\frac{v^2}{r}\mu\left(\frac{a}{a_0}\right)\,,
\end{equation}
where $a\equiv v^2/r$ is the acceleration, $\mu(x)\equiv x/\sqrt{1+x^2}$ is the ``\,standard" interpretation function and $a_0$ is a critical acceleration. Eq.(\ref{eq:finsler1}) is simply the Milgrom's MOND \cite{Milgrom1983a,Milgrom1983b}. According to MOND, the interpolation function $\mu(x)$ can be any positive, smooth and monotonically increasing function which approaches $x$ when $\mid x \mid \ll 1$ and equates to unity when $x$ goes to infinity. If one chooses
\begin{equation}\label{eq:fv}
f^{-1}\big{(}v(r)\big{)}=1+\sqrt{\frac{a_0r^2}{GM}}\,,
\end{equation}
Eq.(\ref{eq:finsler-gravity}) reduces to
\begin{equation}\label{eq:finsler2}
  v^2=\frac{GM}{r}+\sqrt{GMa_0}\,.
\end{equation}
This equation also has a MOND-like behavior with a Bekenstein's ``\,toy" interpolation function $\mu(x)=2x/(1+2x+\sqrt{1+4x}\,)$ \cite{Bekenstein2004}.

The strong gravitational lensing of the early-type galaxies shows that a large number of dark matter is needed in Milgrom's MOND \cite{Ferreras2005,Ferreras2008,Ferreras2009,Ferreras2012}. We try to modify Milgrom's MOND such that the dark matter diminishes. We suppose that the Finslerian rescaling factor is of the form
\begin{equation}\label{eq:fv-quadrupole}
  f^{-1}\big{(}v(r)\big{)}=1+\sqrt{\frac{a_0r^2}{GM}}\left[1+\frac{GMa_0}{v_c^4}\exp\left({-\frac{r}{r_c}}\right)\right]\,,
\end{equation}
where $v_c$ and $r_c$ are two new parameters. The motivation of such a choice comes from the quadrupole effects. The lens galaxy is more likely to be an ellipsoid than a sphere. Thus, quadrupole effects should be taken into consideration. In fact, Milgrom has given a quasi-linear formulation of MOND (QMOND), which also contains the quadrupole contribution \cite{Milgrom2010,Milgrom2012}. Usually, MOND effects vanish in the Newtonian limit. However, Milgrom showed that the quadrupole effects can still exist in high acceleration system. Besides, Li et al showed that the quadrupole effects are essential in explaining the observed convergence $\kappa$-map of the Bullet Cluster 1E 0657\---558 \cite{Li2013}. The last term on the right-hand-side of Eq.(\ref{eq:fv-quadrupole}) is very like the quadrupole term used in reference \cite{Li2013}, except that it is orientation independent now. In other words, it can be regarded as a quadrupole term averaged over the orientation. An exponential cutoff is needed in order to keep the Tully\---Fisher relation. Dropping the last term on the right-hand-side of Eq.(\ref{eq:fv-quadrupole}), the Finslerian MOND reduces to Milgrom's MOND with the Bekenstein's ``\,toy" interpolation function, see Eq.(\ref{eq:fv}). The parameter $v_c$, which has the dimension of velocity, characterizes the strongness of the quadrupole effects. Beyond the scale length $r_c$, the lens galaxy can be seen as spherically symmetric, so that the quadrupole effects vanish.

From the discussion above, we can see that the dynamics in Finsler spacetime is very like that in Minkowski one, except that the spatial distance is rescaled. Following the standard procedure of general relativity, it is easy to obtain the gravitational deflection angle of light in Finslerian MOND \cite{Li2010a}. In the first order approximation, the deflection angle caused by a point source is simply a rescaling of that in general relativity \cite{Li2013}
\begin{equation}\label{eq:deflection-angle1}
  \alpha(b)=\frac{4GM}{bf(v(b))}\,,
\end{equation}
where $M$ is the mass of the source and $b$ is the impact parameter. The generalization to a spherically symmetric mass distribution is straightforward. It reads
\begin{equation}\label{eq:deflection-angle2}
  \alpha(b)=\frac{4Gb}{f(v(b))}\int_0^{\infty}\frac{M(<\sqrt{b^2+z^2})}{[b^2+z^2]^{3/2}}dz\,,
\end{equation}
where $M(<r)$ is the cumulative mass in the sphere of radius $r$. When $f(v(b))=1$, Eq.(\ref{eq:deflection-angle2}) reduces to the result of general relativity \cite{Schneider1992}.

\section{Numerical analysis}\label{sec:lensing}

In this section, we will use the observed lensing data of the early-type galaxies to check the validity of the Finslerian MOND. Our sample lenses are the same as that used in reference \cite{Chiu2011}. This sample is a sub-class of the CASTLES lenses studied in reference \cite{Ferreras2005}, where the authors investigated 18 well-observed lenses with the non-parametric method (the spherical symmetry of the lens is not assumed). Among the 18 lenses, some of them have non-collinear double images, others have more than two images. In this paper, only the double images systems with nearly collinear images are chosen. The shear effects cannot be ignored for non-collinear and quadrupole system. As a final result, our sample consists of ten galaxies. For convenience, the general properties of the lenses are listed in Table \ref{tab:samples}. Column (1) is the standard name of the lensing system. Column (2) and (3) are the observed redshifts of the lens and the source from the CASTLES database\footnote{http://www.cfa.harvard.edu/castles.}, respectively. Column (4) is the projected 2D half-light radius in which contains one half of the total luminosity from reference\cite{Ferreras2005}. Column (5) is the size of the lens, which is roughly the radius of the outmost image from reference\cite{Ferreras2005}.
\begin{table}
\begin{center}
\caption{\small{Properties of the sample lenses. Column (1): the standard name of the lensing system. Column (2): the redshift of the lens. Column (3): the redshift of the source. Column (4): the projected 2D half-light radius of the lens. Column (5): the lens radius which is roughly the outmost of the image.}}
\begin{tabular}[t]{ccccc}
\hline
(1) & (2)& (3) & (4) & (5)\\
 Name   &$z_l$  &$z_s$ &$R_e$[kpc]  &$r_{\rm lens}$[kpc]\\
\hline
Q0142\---100  &0.49  &2.72 &3.1 &5.6\\
HS0818+1227	 &0.39  &3.12 	&4.8   &10.6\\
FBQ0951+2635  &0.24  &1.24   &0.7   &2.1\\
BRI0952\---0115   &0.63  &4.50  &0.5	 &2.5\\
Q1017\---207  &0.78  &2.55   &2.4	 &2.4\\
HE1104\---1805	 &0.73  &2.32   &4.4	 &16.3\\
LBQS1009\---0252  &0.87  &2.74  &1.6  &4.8\\
B1030+074  &0.60  &1.54  &2.6  &4.7\\
SBS1520+530  &0.72  &1.86  &2.2  &6.6\\
HE2049\---2745  &0.50  &2.03  &2.9  &4.1\\
\hline
\end{tabular}\label{tab:samples}
\end{center}
\end{table}

The mass distribution of the lens is often described by the spherically symmetric Hernquist profile, whose 3D density distribution reads \cite{Hernquist1990}
\begin{equation}\label{eq:hernquist}
  \rho(r)=\frac{M}{2\pi}\frac{r_h}{r(r+r_h)^3}\,,
\end{equation}
where $M$ is the total mass of the lens, and $r_h$ is a characteristic radius which is related to the projected 2D half-light radius as $R_e=1.8153r_h$. The integration of Eq.(\ref{eq:hernquist}) gives the cumulative mass profile
\begin{equation}\label{eq:cumulative-mass}
  M(<r)=\frac{Mr^2}{(r+r_h)^2}\,.
\end{equation}
The projection of the Hernquist profile closely approximates the de Vaucouleurs $R^{1/4}$ law observed in most elliptical galaxies, so it is often used to model such galaxies.

The deflection angle of light $\alpha$ is not an observable quantity, it should be related to the image position $\theta$ through the lens equation \cite{Schneider1992},
\begin{equation}\label{eq:lens-equation}
  \beta=\theta-\alpha(\theta,M,b)\frac{D_{ls}}{D_s}\,,
\end{equation}
where $\beta$ is the actual position of the lens, $D_{ls}$ is the angular diameter distance between the lens and the source, while $D_s$ is the angular diameter distance between the observer and the source. They can be calculated through the observed redshifts of the lens and the source. The impact parameter $b$ is related to the lens distance as $b=\theta D_l$. The definition of the diameter distance depends on the cosmological model. As was showed by Ferreras et al. \cite{Ferreras2008}, the cosmological model only slightly affects the results. The following calculations depend on the concordance ${\rm \Lambda}$CDM model of $(\Omega_{\rm m},\Omega_{\rm \Lambda},\Omega_{\rm k})=(0.3,0.7,0)$. The Hubble constant is chosen to be $H_0=72~{\rm km~s}^{-1}~{\rm Mpc}^{-1}$. In order to be in accordance with the rotation curves, the critical acceleration is taken to be $a_0=1.2\times 10^{-10}~{\rm m}~{\rm s}^{-2}$ \cite{Chang2013}. Both the images of the source should be used in order to derive the unknown lens position $\beta$ and lens mass $M$. From Eq.(\ref{eq:lens-equation}), we can see that the lens position $\beta$ is a parameterized function of the lens mass $M$, with the image position $\theta$ as the parameter. There are two images in each system, so Eq.(\ref{eq:lens-equation}) gives two equations. Combining the two equations, we can get expression for $\beta$ and $M$.

With the discussion above, we can now estimate the gravitational mass in the Finslerian MOND. The Finslerian rescaling factor is taken to be of the form of Eq.(\ref{eq:fv-quadrupole}). The two new parameters, $v_c$ and $r_c$, should be fixed in order to calculate the gravitational mass of the lens galaxy. However, we do not know a priori what exact values they should take.  As a reasonable assumption, we set the scale length to be the size of the lens, ie., $r_c=r_{\rm lens}$. From physical considerations, $r_{\rm lens}$ is the lower boundary of $r_c$. Since in the range of $r<r_{\rm lens}$, the spherically symmetric approximation does not hold and the quadrupole effects is not negligible. To derive a reasonable value of $v_c$, we first fix the gravitational mass to be the photometric mass, and solve the lens equation (\ref{eq:lens-equation}) to obtain the values of $v_c$ for each lensing independently. Then, we fix $v_c$ to its average value $\bar{v}_c=143~{\rm km}~{\rm s}^{-1}$ to further estimate the gravitational mass.

\begin{figure}
  \centering
 \includegraphics[width=12 cm]{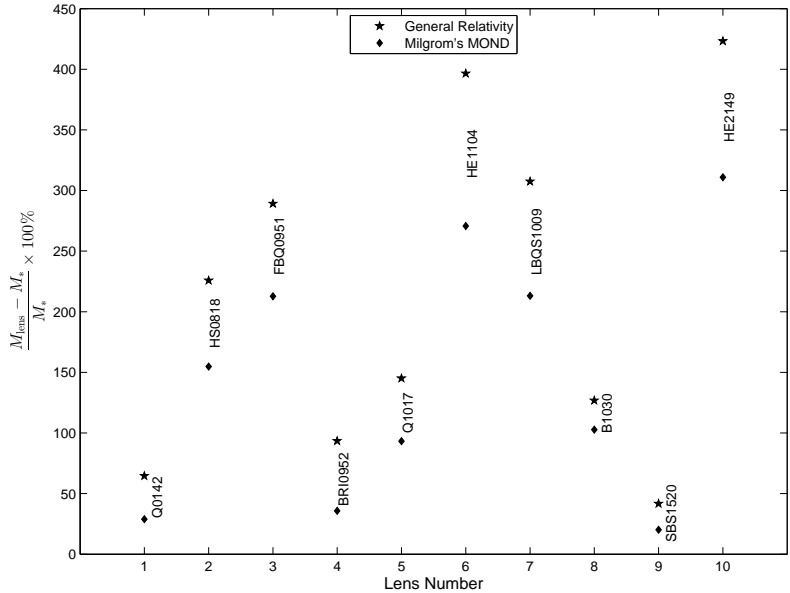}
 \caption{\small{The discrepancy between photometric mass and gravitational mass. The $y$-axis is $(M_{\rm lens}-M_*)/M_*\times 100\%$, where $M_*$ is the photometric mass, and $M_{\rm lens}$ represents the gravitational mass derived from general relativity ($M_{\rm GR}$, pentagram) or from the Milgrom's MOND ($M_{\rm M}$, diamond).}}
  \label{fig:MassDiscrepancy}
\end{figure}

The final results are listed in Table \ref{tab:results}. We also listed the mass derived from general relativity and Milgrom's MOND for comparison. The second column is the photometric mass assuming a Chabrier initial mass function (IMF) cited from reference\cite{Ferreras2005}. The errors are of 90\% confidence level. The third and fourth columns are the masses predicted by general relativity and Milgrom's MOND, respectively. Bekenstein's ``\,toy" interpolation function is used in the Milgrom's MOND. The ``\,standard" interpolation function gives the mass a little larger. The fifth column is the mass derived from the Finslerian MOND. Figure \ref{fig:MassDiscrepancy} shows the discrepancy between the gravitational mass predicted by general relativity or Milgrom's MOND and the photometric mass. As can be seen, the mass required in the Milgrom's MOND is much smaller than that in general relativity. However, the mass estimated in the Milgrom's MOND is still much higher than the photometric mass. It seems that a significant amount of dark matter is still needed in the framework of Milgrom's MOND. Six out of the ten lenses (HS0818+1227, FBQ0951+2635, HE1104\---1805, LBQS1009\---0252, B1030+074 and HE2149\---2745) need more than 100\% dark matter. Especially for HE2149\---2745, about $310\%$ dark matter is required in order to match the observation. Even for the most well fitted lens SBS1520+530, about 20\% mass discrepancy still exists.
\begin{table}
\begin{center}
\caption{\small{The mass estimates in different models. All the masses are in unit of $10^{10}M_{\odot}$. $M_*$ is the photometric mass assuming a Chabrier IMF, with error of 90\% confidence level. $M_{\rm GR}$, $M_{\rm M}$  and  $M_{\rm F}$ are the masses derived from the general relativity, Milgrom's MOND and Finslerian MOND, respectively.}}
\begin{tabular}[t]{ccccc}
\hline
(1) & (2)& (3) & (4) & (5)\\
 Name  &$M_*$   &$M_{\rm GR}$  &$M_{\rm M}$  &$M_{\rm F}$\\
\hline
Q0142\---100  &$20.9_{13.0}^{30.8}$  &34.39  &26.93 &14.16\\
HS0818+1227   &$16.2_{12.6}^{21.2}$	 &52.78  &41.28 	&16.96\\
FBQ0951+2635	&$1.1_{0.5}^{2.1}$  &4.28  &3.44  &2.95\\
BRI0952\---0115	 &$3.5_{2.7}^{4.0}$   &6.77  &4.75  &3.00\\
Q1017\---207	&$4.3_{1.4}^{13.0}$  &10.54  &8.31   &6.05\\
HE1104\---1805	&$22.8_{12.7}^{51.2}$	 &113.20  &84.52   &13.87\\
LBQS1009\---0252  &$5.5_{4.2}^{7.9}$  &22.41  &17.22  &9.11\\
B1030+074  &$10.6_{6.5}^{15.3}$  &24.04  &21.49  &13.96\\
SBS1520+530  &$18.5_{11.2}^{30.9}$  &26.27  &22.22  &10.71\\
HE2149\---2745  &$4.6_{3.6}^{6.7}$  &24.07  &18.90  &10.49\\
\hline
\end{tabular}\label{tab:results}
\end{center}
\end{table}

\begin{figure}
  \centering
 \includegraphics[width=12 cm]{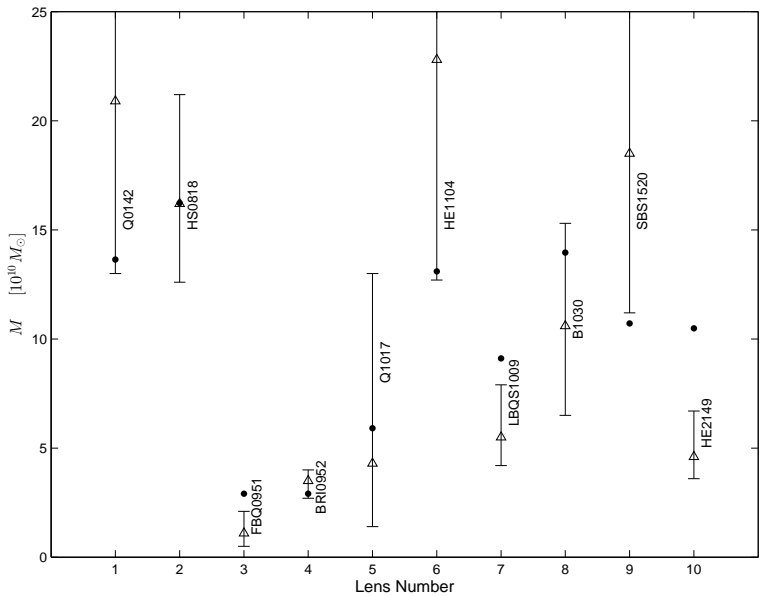}
 \caption{\small{The consistency between photometric mass and gravitational mass. The photometric masses are denoted by triangles with error bars of 90\% confidence level. The black dots are the gravitational masses derived from the Finslerian MOND.}}
  \label{fig:mass}
\end{figure}

From Table \ref{tab:results}, we can see that the gravitational mass in the Finslerian MOND is sharply reduced, much closer to the photometric mass. A more explicit representation of the results is plotted in Figure \ref{fig:mass}. The photometric masses are denoted by triangles, with the error bars of 90\% confidence level. The gravitational masses calculated from the Finslerian MOND are denoted by black dots. As can be seen, for six out of the ten lenses, the gravitational mass matches the photometric mass within errors. For the rest four outliers, three of them deviate from the observation less than $40\%$. The most poorly fitted lens, HE2149\---2745, has mass discrepancy of $\sim 57\%$. The photometric mass, which is deduced from the IMF, has a large uncertainty. The mass cited in our paper is derived from the Chabrier IMF \cite{Chabrier2003}. The Salpeter IMF \cite{Salpeter1955} gives the mass-to-light ratio about 50\% larger than that given by Chabrier IMF. Another uncertainty comes from the photometry observations. The hall-light radius $R_e$ derived from different energy bands may vary. Taking all the uncertainties into consideration, we conclude that dark matter is not indispensable in the framework of Finslerian MOND.

\section{Discussion and conclusions}\label{sec:discussion}

The Finslerian MOND is a generalization of MOND in the Finsler spacetime. The deflection angle of light in the Finsler spacetime is the angle in general relativity multiplied by a Finslerian rescaling factor. By choosing a special rescaling factor, the Finslerian MOND may reduce to the famous MOND of Milgrom. In this paper, we try to explain the mass discrepancy problem of the strong gravitational lensing in the framework of the Finslerian MOND, in absence of dark matter. Assuming an appropriate Finslerian rescaling factor, we find that the gravitational mass is sharply reduced. Although mild deviations exist, there is no strong evidence for the existence of dark matter.

Since there are a lot of uncertainties in the observation, many controversies may arise. Zhao et al. \cite{Zhao2006} analyzed 18 lenses and concluded that MOND can do a successful fit. However, Ferreras et al. \cite{Ferreras2008} investigated six lenses and showed that a significant amount of dark matter is still required  even in the framework of MOND in order to match the observations. One of the reason for the different conclusions is that they got the photometric mass from different methods. The photometric mass in Ref.\cite{Ferreras2008} is about one half of that in Ref.\cite{Zhao2006}. Chiu et al. \cite{Chiu2011} used the same photometric mass as Ref.\cite{Ferreras2008} and found no strong evidence for the existence of dark matter. As they pointed out, it may because Ferreras et al. drew their conclusion by comparing the total lensing mass $M(r<\infty)$ with aperture stellar mass $M(r<r_{\rm lens})$, which led to different results. However, $r_{\rm lens}$ is almost the outmost of the image. Beyond $r_{\rm lens}$, the stellar mass can be omitted. Thus, the aperture stellar mass is a good approximation to the total stellar mass. In our paper, we follow the procedure of Ref.\cite{Ferreras2008} and found that dark matter is not indispensable in the Finslerian MOND.

The Bullet Cluster is a serious challenge to Milgrom's MOND \cite{Angus2006,Angus2007}. The mass profile of Bullet Cluster is neither spherical nor ellipsoidal. A sub-peak appears kpc away from the main peak. For such an asymmetric system, except the quadrupole effects, the dipole contributions should also be accounted. Li et al. have already showed that the Finslerian MOND can well explain the mass discrepancy problem of Bullet Cluster 1E0657-558, by adding the contributions of quadrupole and dipole effects \cite{Li2013}. Besides, Chang et al. used the Finslerian MOND to fit the rotation curves of seventeen spiral galaxies, and they found a universal critical acceleration $a_0\approx1.2\times 10^{-10}~{\rm m}~{\rm s}^{-2}$, which is the same as that of Milgrom's MOND \cite{Chang2013}. It was pointed out that the flatness of the rotation curves may have something to do with the anisotropy of the spacetime.

The Finslerian MOND has some shortcomings. Since we do not know how to write the energy-momentum tensor on the bundle of the manifold, we cannot solve the field equations to derive the Finslerian rescaling factor $f\big{(}v(r)\big{)}$. The factor Eq.(\ref{eq:fv-quadrupole}) we used in this paper is just a hypothesis. The Finslerian MOND has three parameters, ie., the critical acceleration $a_0$, the exponential cutoff length $r_c$ and the characteristic velocity $v_c$. This destroys the simplicity of the model. Of course, one can choose the parameters such that the gravitational mass exactly matches the photometric mass. In order to reduce the freedoms, we fix $a_0$ to the value that obtained from galaxy rotation curves and fix $r_c$ to the size of the lens. The remaining parameter $v_c$, is taken to be the average value of each galaxy. Nevertheless, the Finslerian MOND may be taken as an alternative to the dark matter hypothesis in explaining the rotation curves of spiral galaxies, the weak and strong gravitational lensing of Bullet Clusters, and the strong gravitational lensing of the early-type galaxies.

\begin{acknowledgements}
We are grateful to Y. G. Jiang for useful discussion. This work has been funded in part by the National Natural Science Fund of China under Grant No. 11075166 and No. 11147176.
\end{acknowledgements}


\begin{thebibliography}{}

\bibitem{Milgrom1983a}M. Milgrom, Astrophys. J. {\bf 270}, 365 (1983)

\bibitem{Milgrom1983b}M. Milgrom, Astrophys. J. {\bf 270}, 371 (1983)

\bibitem{Bekenstein2004}J. D. Bekenstein, Phys. Rev. D {\bf 70}, 083509 (2004)

\bibitem{Sanders1996}R. H. Sanders, Astrophys. J. {\bf 473}, 117 (1996)

\bibitem{Sanders:1998gr}R. H. Sanders, M. A. W. Verheijen, Astrophys. J. {\bf 503}, 97 (1998)

\bibitem{Ferreras2005}I. Ferreras, P. Saha, L. L. R. Williams, Astrophys. J. {\bf 623}, L5 (2005)

\bibitem{Ferreras2008}I. Ferreras, M. Sakellariadou, M. F. Yusaf, Phys. Rev. Lett. {\bf 100}, 031302 (2008)

\bibitem{Ferreras2009}I. Ferreras, N. E. Mavromatos, M. Sakellariadou, M. F. Yusaf, Phys. Rev. D {\bf 80}, 103506 (2009)

\bibitem{Ferreras2012}I. Ferreras, N. E. Mavromatos, M. Sakellariadou, M. F. Yusaf, Phys. Rev. D {\bf 86}, 083507 (2012)

\bibitem{Angus2006}G. W. Angus, B. Famaey, H. S. Zhao, Mon. Not. R. Astron. Soc. {\bf 371}, 138 (2006)

\bibitem{Angus2007}G. W. Angus, H. Y. Shan, H. S. Zhao,  B. Famaey, Astrophys. J. {\bf 654}, L13 (2007)

\bibitem{Zhao2006}H. S. Zhao, D. J. Bacon, A. N. Taylor, K. Horne, Mon. Not. R. Astron. Soc. {\bf 368}, 171 (2006)

\bibitem{Clowe2004}D. Clowe, Astrophys. J. {\bf 604}, 596 (2004)

\bibitem{Bradac:2006er}M. Brada\v{c}, D. Clowe, A. H. Gonzalez, P. Marshall, W. Forman, C. Jones, M. Markevitch, S. Randall, T. Schrabback, D. Zaritsky, Astrophys. J. {\bf 652}, 937 (2006)

\bibitem{Brownstein:2007sr}J. R. Brownstein, J. W. Moffat, Mon. Not. R. Astron. Soc. {\bf 382}, 29 (2007)

\bibitem{Adriani2009}O. Adriani, et al., Nature {\bf 458}, 607 (2009)

\bibitem{Aguilar2013}M. Aguilar, et al., Phys. Rev. Lett. {\bf 110}, 141102 (2013)

\bibitem{Li2013}X. Li, M.-H. Li, H.-N. Lin, Z. Chang, Mon. Not. R. Astron. Soc. {\bf 428}, 2939 (2013)

\bibitem{Chang2013}Z. Chang, M.-H. Li, X. Li, H.-N. Lin, S. Wang, Eur. Phys. J. C {\bf 73}, 2447 (2013)

\bibitem{Bao2000}D. Bao, S. S. Chern, Z. Shen, An Introduction to Riemann--Finsler Geometry (Graduate Text in Mathematics, Vol.200). Springer, New York (2000)

\bibitem{Chang2008}Z. Chang, X. Li, Phys. Lett. B {\bf 668}, 453 (2008)

\bibitem{Chang2009}Z. Chang, X. Li, Phys. Lett. B {\bf 676}, 173 (2009)

\bibitem{Li2010a}X. Li, Z. Chang, Phys. Rev. D {\bf 82}, 124009 (2010)

\bibitem{Li2010b}X. Li, Z. Chang, Chinese Phys. C {\bf 34}, 28 (2010)

\bibitem{Milgrom2010}M. Milgrom, Mon. Not. R. Astron. Soc. {\bf 403}, 886 (2010)

\bibitem{Milgrom2012}M. Milgrom, Mon. Not. R. Astron. Soc. {\bf 426}, 673 (2012)

\bibitem{Schneider1992}P. Schneider, J. Ehlers, E. E. Falco, Gravitational Lenses. Springer-Verlag, Berlin (1992)

\bibitem{Chiu2011}M.-C. Chiu, C.-M. Ko, Y. Tian, H. S. Zhao, Phys. Rev. D {\bf 83}, 063523 (2011)

\bibitem{Hernquist1990}L. Hernquist, Astrophys. J. {\bf 356}, 359 (1990)

\bibitem{Chabrier2003}G. Chabrier, PASP {\bf 115}, 763 (2003)

\bibitem{Salpeter1955}E. E. Salpeter, Astrophys. J. {\bf 121}, 161 (1955)

\end{thebibliography}


\end{document}